\documentclass[dvips]{article}

\usepackage{icrctc07}

\newcommand{\degree}{\ensuremath{^\circ}}

\title{Observations of the Crab Nebula and Pulsar with VERITAS}

\shorttitle{Observations of the Crab with VERITAS}

\authors{O. Celik$^{1}$, for the VERITAS Collaboration$^{2}$}
\shortauthors{O. Celik et al.}
\afiliations{$^1$University of California, Los Angeles, $^2$For full author list see G. Maier, ``VERITAS: Status and Latest Results", 
these proceedings}
\email{celik@astro.ucla.edu}

\abstract{Observations of the Crab Nebula have proven to be the best tool to calibrate and to characterize the performance of a Cherenkov telescope. Scientifically, it is interesting to measure the energy spectrum of the Crab Nebula close to the inverse-Compton peak where a deviation is expected from the power law seen at energies above 300 GeV. Additionally, it is important to search for pulsed emission from the Crab Pulsar at energies beyond the 10 GeV upper limit of the EGRET pulsar detection. Since current models predict a cut-off in pulsed emission between 10 and 100 GeV, measurements at energies close to this range may help to discriminate between them. We observed the Crab extensively in the 2006-2007 season during the VERITAS 2- and 3-telescope commissioning phases. Using this data set we reconstructed a preliminary energy spectrum of the signal from the Crab Nebula. We also measured the optical pulsed signal to validate our GPS time-stamping and barycentering techniques and obtained an upper limit for the pulsed emission at gamma-ray energies.}

\begin{document}
\maketitle

%Begin the section.
\section{Introduction}
VERITAS, located in southern Arizona (32\degree N, 111\degree W) at an elevation of 1268 m, is an observatory designed to explore the very high energy (VHE) gamma-ray sky in the energy band between 100 GeV and 50 TeV \cite{Gernot}. It consists of four 12m diameter Cherenkov telescopes, each with a 499-pixel camera. It uses the stereoscopic imaging atmospheric Cherenkov technique (IACT) to detect gamma rays from astronomical sources. VERITAS has moved through phases of 2-, 3- and 4-telescope operation during its construction and commissioning this past year.  We observed the Crab Nebula and Pulsar extensively during the 2- and 3-telescope phases. This paper discusses the analysis of these data. We also present a preliminary differential energy spectrum of the nebula and a preliminary upper limit for the pulsed gamma-ray emission.

\section{The Crab Nebula and Pulsar}
The Crab Nebula is a plerion-type supernova remnant (SNR). It was the first source definitively detected at TeV energies, by the Whipple collaboration \cite{WhDet}. Its spectral energy distribution is well measured across a broad range of wavelengths and it is considered the standard candle in TeV astronomy due to its strong, stable emission. The nebula emission is believed to be powered by the conversion of the rotational slowdown of the central pulsar to synchrotron radiation. Inverse-Compton scattering of those synchrotron photons with the primary accelerated electrons gives rise to the emission in the gamma-ray regime as explained by the Synchrotron-Self-Compton model \cite{SSC}. The Crab Nebula spectrum has been measured independently by many groups above $\sim$200 GeV, with good agreement \cite{WhSpect, HGSpect, MGSpect, HSSpect}. Consistency with this established spectrum is an important test for any new gamma-ray telescope.

At the center of the Crab Nebula lies the Crab Pulsar (PSR 0531 +21). Periodic emission from the pulsar, with a 33 ms period, has been observed in all energy regimes with the exception of very high energy gamma rays. The EGRET detection of pulsed emission in the 100 MeV - 10 GeV energy band, with no evident turnover of its spectrum, is currently the highest energy detection \cite{EGRET}. None of the ground-based gamma-ray experiments sensitive in the adjacent VHE domain has yet provided any detections. Currently there are two primary models attributing the pulsed emission to curvature and synchrotron radiation from the energetic electrons/positrons accelerated in the strong magnetic fields of the pulsar magnetosphere. These models differ by whether the emission region is near to the magnetic poles of the neutron star (the polar cap model \cite{ref4}), or near to the light-cylinder at the outer magnetosphere (the outer gap model \cite{ref5}). 
In the polar cap model, a super-exponential cut-off occurs in the pulsed spectrum at a few GeV due to the magnetic pair production in the strong field close to the pole. On the other hand, in the outer-gap model a cut-off at a few tens of GeV is expected due to photon-photon pair production, which has a weaker energy dependence. By detecting the pulsed emission or placing stringent upper limits in the GeV-TeV energy range, one can potentially discriminate between these models. VERITAS has a better sensitivity at lower energies than previous-generation IACT instruments. This suggests the possibility of detecting the pulsed emission, if an outer gap model applies, or of strongly constraining both models if pulsed emission is not detected.

\section{Data Acquisition and Analysis}
The data set used in this analysis was selected from the observations made using 2 and 3 telescopes between October 2006 and February 2007. During that time, different trigger and observing configurations were tested to determine the optimum configuration for the array. Only data with this optimum configuration are used in the present analysis. 
Data-quality selection, based on the stability of trigger rates, response of the instrument to background cosmic rays and the deadtime of the system, was additionally used to reject low-quality data. The final data set, after this quality selection, is 15.7 hours of 2-telescope data and 3.3 hours of 3-telescope data. Data on both sets were collected in the so-called ``wobble'' observing mode, for which the telescopes are pointed such that the source is at a 0.5\degree offset from the position of the camera center. For 2-telescope data we required that both telescopes trigger within a certain coincidence time, while for 3-telescope data we required that any two out of three telescopes trigger to record the event.

%\section{Analysis}
The data were analyzed using the standard \mbox{VERITAS} analysis package 
\cite{Michael, Peter} in which calibration, cleaning and parametrization of the signal, reconstruction of the shower parameters and the gamma/hadron separation cuts are applied to the data. 
%The cuts used in this analysis are listed in Table \ref{cuts}. 
The quality-selection cuts which require the total recorded signal ($Size$) to be greater than 400 d.c, the number of pixels in the image ($\# ofPix$) to be greater than 5 and the distance of its centroid from the center of the camera ($Dist.$) to be in the range (0.05\degree , 1.3\degree) are applied before the stereo reconstruction. The gamma/hadron separation cuts based on the mean scaled width and length ($0.05<MSW<0.95$ and $0.05<MSL<1.15$ respectively) and the squared distance between the source and the reconstructed gamma ray position ($\theta^2<0.025$) are applied at the final step. 
The same data were also analyzed using an independent analysis package to confirm the results. 

%\begin{table}[t]
%\begin{center}
%\begin{tabular}{|c|c|}
%\hline
%Quality Selection                  & Gamma/Hadron Separation \\
%\hline
%     $Size > 400~dc.$              & $0.05 < MSW < 0.95$ \\
%     $\# ofPix > 5$                & $0.05 < MSL < 1.15$  \\
%$0.05\degree < Dist. < 1.3\degree$ & $\theta^2 < 0.025$  \\
%\hline
%\end{tabular}
%\caption{ The standard cuts used in the analysis. The various quantities are defined in the text.} \label{cuts}
%\end{center}
%\end{table}

\subsection{Detection}
The Crab Nebula is strongly detected during both 2-telescope and 3-telescope phases. The detection is at the $7~\gamma/\textrm{min}$ level with a significance of $31~\sigma/\sqrt{\textrm{hr}}$ for three telescopes and at $3.4~\gamma / \textrm{min}$ level with a significance of $20~\sigma/\sqrt{\textrm{hr}}$ for two telescopes using the cuts described above.
%listed on Table \ref{cuts}. 
%In the upper panel of Figure~\ref{thSq}, the $\theta^2$ distributions for the 2-telescope data (left) and the 3-telescope data (right) are shown. In the lower panel a sum of two Gaussian functions \cite{HSSpect} is fit to the excess $\theta^2$ distribution for each data set. 
The excess $\theta^2$ distributions for each data set can be described by a sum of two Gaussian functions \cite{HSSpect}.  
The fit to 
3-telescope data gives $A_{Rel}=0.17$ with $\sigma_{1}=0.067$ and $\sigma_{2}=0.13$. 

%\begin{figure}[htpb]
%\begin{center}
%\noindent
%\includegraphics[width=68mm]{icrc1209_fig1.eps}
%\end{center}
%\caption{Theta-square distribution (upper) and its fit to a sum of two Gaussian functions (lower) for 2-telescope data (left) and 3-telescope data (right)}\label{thSq}
%\end{figure}

\subsection{Energy Spectrum}
\begin{figure}[htpb]
\begin{center}
\noindent
\includegraphics[width=50mm]{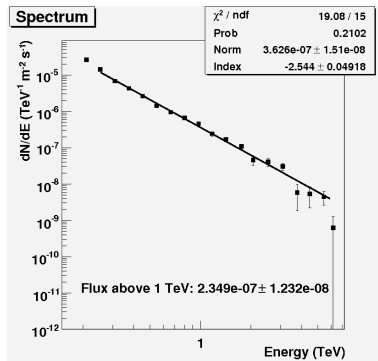}
\end{center}
\caption{The reconstructed Crab Energy Spectrum using 3-telescope data set.}\label{Spectrum}
\end{figure}

The energy for each succesfully reconstructed event is estimated using lookup tables based on the $Size$, the impact parameter of the event and the zenith angle. The lookup tables are generated using the simulations of the gamma-ray showers and the detailed detector response. 
%The bias in the estimated energy defined as $(E_{reco} - E_{true})/E_{true}$ is shown for different zenith angles on the left panel of the Figure~\ref{Energy}. 
Effective collection area of the instrument as a function of energy of the event, the zenith angle and offset of the source from the pointing direction is also determined using the same simulations.
% (Figure~\ref{Energy}). 
To calculate the differential flux from the Crab Nebula, a looser gamma/hadron seperation cuts are applied to the data to keep more events in lower energy bins. In each energy bin, the gamma-ray events passing the cuts are weighted by the inverse of the effective area of the instrument for the energy of the event, and the difference of this weighted sum of the gamma-ray events from the on source region and the normalized weighted sum of the gamma-ray events from the background regions is calculated. The differential flux for each energy bin is determined by normalizing this difference by the live-time for that energy bin and the bin width. Figure~\ref{Spectrum} shows the measured energy spectrum of the Crab Nebula using the 3-telescope data set fitted to a simple power law between 250 GeV and 6 TeV. The minimum energy bin in the fit is defined by requiring the energy bias lower than 10\% and the maximum energy bin is defined by requiring a detection of minimum of 2 sigma in each bin. The spectrum is checked for consistency using two independent analysis packages and it also agrees very well with results from other experiments.

%\begin{figure}[htpb]
%\begin{center}
%\noindent
%\includegraphics[width=68mm]{icrc1209_fig1.eps}
%\end{center}
%\caption{Left: The energy bias, as defined in the text, for zenith angles 0\degree (black circles), 20\degree (red squares), 30\degree (green triangles) and 40\degree (blue inverted triangles). The vertical lines show the energy thresholds for 10\% bias for each zenith angle. Right: The effective collection area of the detector for the same zenith angles as in figure on the left.}\label{Energy}
%\end{figure}

\section{Temporal Analysis}
The temporal analysis of the Crab signal is used to search for pulsed emission from the Crab Pulsar. The event arrival times are tagged online by GPS clocks at each telescope 
%with precision of 100 ns. The GPS time of each event is 
and are transformed to the solar system barycenter to compensate for the motion of Earth. We then calculate the relative phase of the pulsar rotation for each event with respect to a reference epoch 
%according to following equation. 
%\[ \phi(T) = \phi(T_{0}) + f(T_0)(T - T_0) + {\frac{1}{2}} \dot{f} (T_0)(T - T_0)^2 \]
%The reference epoch is taken as 
which is the arrival time of the first radio pulse at the solar system barycenter. The radio ephemerides for the Crab pulsar are obtained from the Jodrell Bank Observatory \cite{ref9}. Barycentering libraries used in this analysis were developed \cite{ref10} and tested for this analysis using the optical data taken with a special set-up explained in the following section. The same data set is also tested for periodicity by an independent temporal analysis \cite{ref11}.  
   
\subsection{The Crab Optical Pulsar}
\begin{figure}[htpb]
\begin{center}
\noindent
\includegraphics[width=50mm]{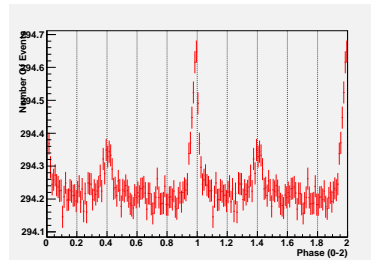}
\end{center}
\caption{The phase histogram of the optical signal detected by VERITAS.}\label{optSig}
\end{figure}

Before analyzing the gamma-ray data for periodicity, we took special runs using only the central pixel of one telescope to detect the optical pulsed signal from the Crab pulsar. This detection of the periodic optical signal confirms that our online time stamping and our offline data analysis are capable of detecting the 33 ms pulsed emission.
%and that we can reconstruct the double-peaked phase structure of the Crab pulsar
For the special optical runs, we ran the system at the single photo-electron trigger level, triggering the system at 20 kHz and recording only the time-stamps and number of pulses collected between each trigger from the central pixel and a pixel at the outer edge of the camera. Figure~\ref{optSig} shows the phase plot of the signal obtained by combining the data from 3 runs taken on 2 different nights, for a total of 30 minutes. It clearly shows the main and inter pulses of the optical pulsed emission.  

\subsection{The Pulsed Emission Upper Limit}
\begin{figure}[htpb]
\begin{center}
\noindent
\includegraphics[width=68mm]{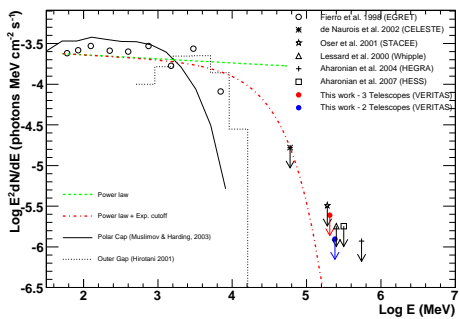}
\end{center}
\caption{The integral flux upper limits for the pulsed emission obtained from 3-telescope and 2-telescope data set are shown with filled circles along with the results from other experiments.}\label{UL}
\end{figure}

Since the pulsed emission is expected only at the lowest energies to which VERITAS is sensitive, the $Size$-cut used in the standard VERITAS analysis 
%(Table \ref{cuts}) 
removes all relevant events. Thus, to search for the pulsed emission, all data which pass the standard quality cuts with the exception of the $Size$-cut are used. Since the angular resolution is degraded at low energy, the standard $\theta^2$-cut was also relaxed to $\theta^2 < 0.04$ to enhance the acceptance of low-energy events. The binning of events by phase acts as a separation of signal and background, and no further gamma/hadron separation cuts are applied. These loose cuts increased the overall number of candidate gamma-ray events by a factor of 27. 
%In Figure~\ref{gmmPlsr} the phase histogram of the signal from the source region and the estimated background is shown for the 2-telescope data (left panels) and 3-telescope data (right panels) with the loose cuts explained in this section (upper panels) and with the standard cuts (lower panels). 
The significance of the pulsed excess is calculated by assuming the pulse shape as seen in the EGRET energy range, with a defined pulsed emission region between phases $[-0.06-0.04]$ and $[0.32-0.43]$ for the main pulse and inter-pulse regions respectively. The remaining phase regions are used to estimate the background for the pulsed emission. No evidence for pulsed emission is seen. An upper limit for the pulsed emission is calculated using the method of Helene \cite{Helene}, assuming Gaussian statistics. The upper limit on the integral flux of the pulsed signal is then derived separately for the 3-telescope and 2-telescope data. The integral flux upper limit obtained using 3-telescope data set is $8.17 \times 10^{-12} cm^{-2}s^{-1}$ above the energy threshold of 200 GeV and the upper limit for 2-telescope data set is $3.5 \times 10^{-12} cm^{-2}s^{-1}$ above 240 GeV as plotted on figure~\ref{UL}.

\section{Conclusion}
Steady emission from the Crab Nebula is detected at very high significance by VERITAS. The steady signal from the Crab was used to perform various performance checks on the system for different configurations. A preliminary differential spectrum of the Crab was measured. A simple power law fit to the data points gives $\frac{dN}{dE} = (3.63 \pm 0.15) \times 10^{-7} (\frac{E}{1 TeV})^{(-2.54 \pm 0.05)}~TeV^{-1} m^{-2} s^{-1}$. Pulsed emission from the Crab Pulsar is also searched for with the same data set, using looser cuts. Integral flux upper limit for the pulsed emission of $8.17 \times 10^{-12} cm^{-2}s^{-1}$ above the energy threshold of 200 GeV is calculated from the 3-telescope data set.

\subsection*{Acknowledgments}
This research is supported by grants from the U.S. Department of Energy,
the U.S. National Science Foundation,
and the Smithsonian Institution, by NSERC in Canada, by PPARC in the UK and
by Science Foundation Ireland.

%This is the reference to .bib file (Without .bib!)
\bibliography{icrc1209}
%This in the bibtex style, is ok.
\bibliographystyle{plain}
\end{document}